# Dynamics of Natural Killer cell receptor revealed by quantitative analysis of photoswitchable protein

Sophie V. Pageon*[1,2], Gerardo Aquino*[1], Kathryn Lagrue[1,3], Karsten Köhler[1], Robert G. Endres[1,4] and Daniel M. Davis[1,3,4]

[1]Department of Life Sciences, Imperial College London, Exhibition Road, London, SW7 2AZ, UK.

[2]Present address: Centre for Vascular Research, Lowy Cancer Research Centre, University of New South Wales, Kensington, New South Wales 2052, Australia.

[3]Manchester Collaborative Centre for Inflammation Research, Core Technology Facility, University of Manchester, Oxford Road, Manchester, M13 9PT, UK.

[4]Correspondence to: Daniel M. Davis, E-mail: daniel.davis@manchester.ac.uk or Robert G. Endres, E-mail: r.endres@imperial.ac.uk

* Sophie V. Pageon and Gerardo Aquino contributed equally to this work.

**Abstract**

Natural Killer (NK) cell activation is dynamically regulated by numerous activating and inhibitory surface receptors that accumulate at the immune synapse. Quantitative analysis of receptor dynamics has been limited by methodologies which rely on indirect measurements such as fluorescence recovery after photobleaching. Here, we report a novel approach to study how proteins traffic to and from the immune synapse using NK cell receptors tagged with the photoswitchable fluorescent protein tdEosFP, which can be irreversibly photoswitched from a green to red fluorescent state by ultraviolet light. Thus, following a localized switching event, the movement of the photoswitched molecules can be temporally and spatially resolved by monitoring fluorescence in two regions of interest. By comparing images with mathematical models, we evaluated the diffusion coefficient of the receptor KIR2DL1 ($0.23 \pm 0.06$ $\mu m^2 s^{-1}$) and assessed how synapse formation affects receptor dynamics. Our data conclude that the inhibitory NK cell receptor KIR2DL1 is continually trafficked into the synapse and remains surprisingly stable there. Unexpectedly however, in NK cells forming synapses with multiple target cells simultaneously, KIR2DL1 at one synapse can relocate to another synapse. Thus, our results reveal a previously undetected inter-synaptic exchange of protein.

**Introduction**

NK cells are characterized by their ability to lyse a variety of diseased cells without the need for prior antigen recognition. One way in which NK cells can discriminate healthy cells from abnormal cells is, according to the 'missing-self' hypothesis, that NK cells can recognize and eliminate target cells which have down-regulated surface expression of class I major histocompatibility complex (MHC) proteins, a feature of many virally-infected or transformed cells (1, 2). Broadly, NK cell activation is regulated by the balance of signals from a large repertoire of activating and inhibitory receptors expressed at their surface. When an NK cell meets a potential target cell, activating receptors can be ligated. If the appropriate MHC class I protein is also present on the target cell, inhibitory receptors will be engaged to transduce inhibitory signals that can interfere with activating signaling to protect the target cell from being lysed. Inhibitory receptors expressed by human NK cells include three main families: killer immunoglobulin-like receptors (KIRs), the CD94/NKG2 receptors, and the leukocyte immunoglobulin-like receptors (LIRs, also known as ILT). Receptors from the KIR family recognize classical MHC class I proteins - human leukocyte antigens (HLA)-A, -B and -C (3–5).

Imaging NK cells in contact with susceptible and resistant target cells has shown that inhibitory signals do not inhibit cell function globally, but are spatially restricted towards a specific target cell (6, 7). This is facilitated by the assembly of an immune synapse - a cell-cell interface at which receptors and signaling molecules organize. The formation of an activating or cytolytic synapse is dependent on ATP and the reorganization of the actin cytoskeleton (8, 9). However, the accumulation of inhibitory receptors and ligands at the NK cell synapse has been shown to occur, to a large extent, independently from remodeling of the actin cytoskeleton or other ATP-driven processes (6, 10–12). Importantly however, there has been a long-standing controversy as to whether or not the accumulation of KIR receptors at synapses is entirely independent of the cytoskeleton, since the speed of KIR accumulation at the synapse is influenced by pharmacological inhibitors of actin polymerization (11). However, diffusion of KIR proteins in the NK cell membrane and the dynamics of the recruitment of KIR to the immune synapse have been little studied quantitatively.

More broadly, cellular signaling and cell-fate decisions are tightly regulated and dependent on protein dynamics which are difficult to observe in real time. Typically, imaging of protein behavior in cell membranes has relied on indirect measurements based on photobleaching of fluorophores. However, these methods do not allow direct visualization of proteins within cells. Here, to demonstrate our quantitative analysis of photoswitchable fluorescent proteins for studying protein dynamics, we consider the specific example of receptor recruitment to an immune synapse. By photoconverting molecules from green to red fluorescence in selected regions of the cell membrane, this technique can be used to monitor the movements of KIR2DL1 within NK cell membranes. We found that in unconjugated NK cells, the dynamics of KIR2DL1 fit a free diffusion model. In the presence of a target cell expressing an appropriate ligand for KIR2DL1, the KIR2DL1 protein was continuously recruited to the immune synapse, where it remained stable, unless a second synapse was formed, in which case KIR2DL1 was able to traffic from one synapse to another. We explored the mechanism of accumulation within the synapse, using different mathematical models of KIR dynamics and found that receptor confinement in the synapse can be modeled by imposing either an apparent 'force' or reduced diffusion in the synapse.

**Materials and Methods**

*Cell culture.* The NK cell line YTS and the target cell line 721.221 (an EBV-transformed B cell line selected to lack endogenous cell surface expression of HLA proteins; hereafter referred to as 221) were maintained in RPMI-1640 (GIBCO) supplemented with 10% FCS, 2mM L-glutamine, and 1mM penicillin/streptomycin (all Invitrogen; complete medium). 221 cells transfected to express HLA-Cw4 (221/Cw4) were generated as previously described (13). YTS cells transfected to express tdEosFP-tagged KIR2DL1 (YTS/KIR2DL1-tdEosFP) were generated by retroviral transduction of YTS cells as previously described (14). Briefly, the packaging cell line Phoenix-amphotropic (Nolan lab, Stanford) was transfected with the retroviral vector PINCO encoding the KIR2DL1-tdEosFP fusion protein using Lipofectamine LTX (Invitrogen). Viral supernatant collected 24 and 48 hours post-transfection was used in three sequential centrifugations (45 minutes, 1600 rpm, 30°C) for infection of $10^6$ YTS cells. Two weeks post-infection, cells expressing KIR2DL1 were enriched by flow cytometry.

*Plasmid generation.* We have described the generation of the KIR2DL1-tdEosFP construct elsewhere (14). In brief, to generate the C-terminal tag of KIR2DL1*001 with tdEosFP, the coding sequence for tdEosFP (Mobitec, Germany) was inserted to replace the GFP in the retroviral vector PINCO expressing KIR2DL1-GFP (15) using the *Bam*HI/*Not*I sites.

*Live cell confocal fluorescence microscopy.* For live cell microscopy, cells were imaged in eight-well chambered borosilicate coverglasses (LabTek, Nunc, Rochester, NY) pre-coated with 10 μg ml$^{-1}$ fibronectin (Sigma). Cells were imaged by confocal microscopy using an SP5 inverted confocal microscope (TCS SP5 RS, Leica Microsystems) using a 63x water immersion objective with a numerical aperture of 1.2. Simultaneous imaging in different channels was performed by sequential line scanning. Live cell images were acquired at 37°C and 5% $CO_2$. YTS/KIR2DL1-tdEosFP and 221/Cw4 cells were gently added to fibronectin-coated chamber slides at a 1:1 ratio just before imaging. During live cell acquisitions, a small region of the YTS/KIR2DL1-tdEosFP membrane was selected and the tdEosFP molecules in this region were photoconverted from green to red fluorescence (FRAP Wizard, Leica). This photoswitching involved firing the 405 nm UV laser (at 80% laser power) at a specific region of the cell membrane for seven seconds. The cells were then imaged for a further 20 minutes at one frame every ten seconds to follow the movement of the photoswitched molecules. The green emission from tdEosFP was visualized using the 496 nm laser line (fluorescence was collected between 500 and 550 nm) and the red emission wavelengths were visualized using excitation with a 543 nm laser line, and fluorescence collected at 575-650 nm.

*Image processing.* Imaging data obtained from the photoswitching experiments were processed and analyzed (ImageJ, US National Institutes of Health, Bethesda, MA). To correct for movement of cells or cell conjugates across the field of view during acquisition, the frames within each image sequence were aligned (plug-in for ImageJ, "StackReg", Philippe Thévenaz, Swiss Federal Institute of Technology Lausanne, Switzerland). The images were then assigned a threshold to minimize background fluorescence before further analysis. Measurements in the red channel were taken at the region where tdEosFP molecules were photoconverted to red during the acquisition as well as the membrane region directly opposite (Fig. S1 in the Supporting Material). Fluorescence was monitored in narrow strip regions to take into account any remaining cell movement and changes in cell shape that might have occurred during the acquisition. For each region, the number of fluorescent pixels within that region was measured. This method was validated by measuring the length of membrane detected within the two monitoring regions in a typical image sequence (Fig. S1). The error of the mean was smaller than 0.6%, indicating that for each region that is selected, the length of membrane under analysis remains constant for the whole image sequence. Following validation, this method was applied to all cells in further experiments. The data were normalized to the first time point after photoswitching. This normalization allowed the comparison of different cells that might have had different initial levels of red fluorescence after photoswitching.

*Mathematical modeling.* Given the size of the initial photoswitched area, an accurate estimate of the receptor density at any subsequent time can be determined by modeling receptor diffusion over a surface. When no target cell is attached to the NK cell, i.e. when no immune synapse is formed, a reasonable assumption is that of free diffusion of the receptor on a membrane. The distribution of the receptors is then obtained by exactly solving the free-diffusion equation on a spherical cell:

$$\frac{\partial p(r,\theta,\varphi,t)}{\partial t} = D\nabla^2 p(r,\theta,\varphi,t) \qquad (1)$$

where $p(r,\theta,\varphi,t)$ is the receptor density on a point on the sphere of radius $r$ at a given azimuthal angle $\theta$ and polar angle $\varphi$. $D$ is the diffusion coefficient, which is directly proportional to the receptor mobility through the Einstein relation, and $\nabla^2$ the Laplacian differential operator, which on a sphere of radius $r$ is given by:

$$\nabla^2 = \frac{1}{r^2 \sin\theta^2}\frac{\partial^2}{\partial\varphi^2} + \frac{1}{r^2 \sin\theta^2}\frac{\partial}{\partial\theta}\left(\sin\theta^2 \frac{\partial}{\partial\theta}\right). \qquad (2)$$

The general solution of the free-diffusion equation is:

$$p(r,\theta,\varphi,t) = -2\pi \sin\theta_0 \, P^0{}_l(\cos\theta_0) + \sum_{l=1}^{L} \frac{\sin\theta_0 \, P^1{}_l(\cos\theta_0)}{2l(l+1)(2l+1)} e^{-Dl(l+1)t}, \qquad (3)$$

which depends only on $D$ and the initial condition $\vartheta_0$. $P^1{}_l(x)$ are the Legendre polynomials. The exact solution is obtained for $L = \infty$ but convergence is obtained for $L \gtrsim 20$ terms in the sum. The initial condition is given by the experimental procedure and therefore corresponds to the projection on the sphere of a rectangular strip of size $s_0 = 10\mu m$ x $1.4\ \mu m$, i.e. the area activated through controlled photoswitching. The above solution, integrated over the experimental detection area, is proportional to the recorded red light intensity coming from that area. As dictated by the experimental protocol, this area is taken either within or opposite to the photoswitched area. The diffusion coefficient and proportionality constant (converting the integrated density of receptors into their red luminosity) were fitted using minimization of $\chi^2$ quality-of-fit function.

In the presence of a target cell, imaging data show an accumulation of receptors at the synaptic attachment with the target cell, ruling out that free diffusion is taking place inside the synapse. We therefore resort to numerical modeling, adding to free diffusion inside the synapse the binding and unbinding process of KIR2DL1 receptors to ligands on the target cell and either of the two following conditions: (i) recruitment of KIR2DL1 receptors modeled with an attractive force, which traps the receptors in the synapse, or (ii) a slower diffusion coefficient due to trafficking occurring only inside the synapse, effectively trapping the receptors therein.

Both models include free diffusion outside the synapse, i.e. Eq. 1. Inside the synapse, model (i) is mathematically described by the following equations:

$$\frac{\partial p_u(r,\theta,\varphi,t)}{\partial t} = -k_b\sigma(r,\theta,\varphi,t)p_u(r,\theta,\varphi,t) + D\nabla^2 p_u(r,\theta,\varphi,t) + k_u p_b(r,\theta,\varphi,t)$$
$$+ K_S(\nabla \cdot F)p_u(r,\theta,\varphi,t) \quad (4a)$$

$$\frac{\partial p_b(r,\theta,\varphi,t)}{\partial t} = -k_u p_b(r,\theta,\varphi,t) + k_b\sigma(r,\theta,\varphi,t)p_u(r,\theta,\varphi,t), \quad (4b)$$

with $F$ the force field in the synapse, of strength $K_S$, $\sigma(r,\theta,\varphi,t)$ the density of ligand on the target cell and $p(r,\theta,\varphi,t) = p_b(r,\theta,\varphi,t) + p_u(r,\theta,\varphi,t)$, the total density of receptors in the synapse, given by the sum of those bound to the target cell $p_b$ and those unbound $p_u$. Eqs. 4 are of reaction-diffusion type with a force field, and can be interpreted as a Fokker-Planck equation (16) with the addition of the terms carrying the parameters $k_b$ and $k_u$, which describe the binding and unbinding of receptors to ligand on the target cell inside the synapse. Setting $K_S=0$ in Eqs. 4 amounts to having only uniform diffusion of receptors with binding and unbinding to ligand inside the synapse.

In contrast, model (ii) describes the receptor dynamics inside the synapse as follows:
$$\frac{\partial p_u(r,\theta,\varphi,t)}{\partial t} = -k_b\sigma(r,\theta,\varphi,t)p_u(r,\theta,\varphi,t) + D_S\nabla^2 p_u(r,\theta,\varphi,t) + k_u p_b(r,\theta,\varphi,t) \quad (5a)$$
$$\frac{\partial p_b(r,\theta,\varphi,t)}{\partial t} = k_b\sigma(r,\theta,\varphi,t)p_u(r,\theta,\varphi,t) - k_u p_b(r,\theta,\varphi,t), \quad (5b)$$

with $D_S < D$ setting slow diffusion inside the synapse. Notice that according to both models only unbound receptors can diffuse inside the synapse, while bound receptors are immobile (no diffusive term in Eqs. 4b and 5b). For the binding and unbinding rates $k_b$ and $k_u$ we adopt throughout the paper the values $2\times10^5$ M$^{-1}$s$^{-1}$ and 2 s$^{-1}$, as derived from experiments (17, 18).

We numerically integrated the above equations on a planar grid with periodic boundary conditions, using once more a small area of size $s_0$ either opposite to the synapse or within the synapse. The models were subsequently fitted to the fluorescence data from the experiments. The fitting parameters were the diffusion coefficient and the force parameter for the force model and the two diffusion coefficients (inside and outside the synapse) for the slow diffusion model. An overall rescaling parameter was also used to account for the conversion of the integrated receptor density provided by the models into light intensity. This rescaling parameter was chosen differently for the conversion at the synapse and the conversion opposite to the synapse, since the presence of the target cell at the synapse is expected to affect the transmitted fluorescence.

## Results

**Mobility of KIR2DL1 in unconjugated NK cells**

To investigate the membrane dynamics of the inhibitory receptor KIR2DL1, the immortal NK cell line YTS was transfected to express KIR2DL1 fused to the photoswitchable protein tdEosFP. A stable cell line expressing KIR2DL1-tdEosFP was established, as demonstrated by flow cytometry and confocal microscopy, and the functionality of the construct in these cells was tested in cytotoxicity assays (14). The photoswitchable fluorescent protein tdEosFP is capable of an irreversible photoconversion from green fluorescence to red fluorescence upon near-UV irradiation at ≈390 nm (19–21). With the development of such proteins, it has become possible to selectively photoactivate fluorescent molecules in specific regions of the cell membrane by locally irradiating these regions with a short pulse of near-UV light, as demonstrated here in Fig. 1.

To investigate the diffusion of KIR2DL1 in the membrane of unconjugated NK cells, YTS/KIR2DL1-tdEosFP cells were imaged by confocal microscopy in the absence of target cells. A small portion of the membrane was selected and tdEosFP molecules were photoconverted from green to red fluorescence within this region by illumination with the UV laser. Each cell was imaged for 20 minutes after photoswitching to follow the movement of the KIR2DL1-tdEosFP fusion protein in the membrane of the cell, during which time the red fluorescence was seen to diffuse around the cell, until it was uniformly distributed in the cell membrane (Fig. 2A). For each cell, the fluorescence both at the site of photoswitching and in the region opposite the site of photoswitching was quantified, analyzed and modeled (Fig. 2B, C & D). An exponential decrease in the amount of red fluorescence at the site of photoswitching was observed, coupled with a gradual increase of red fluorescence in the region opposite the site of photoswitching (Fig. 2D). Importantly, the red fluorescence within the cell membrane appears to reach equilibrium in the two sites of measurement, as demonstrated by the two curves meeting on the graph after approximately 1000 seconds (16-17 minutes).

By fitting the free diffusion model in Eq. 3 to the mean intensity data obtained from experiments on $n = 10$ cells, with deviations weighted by the inverse of the experimental variances, the diffusion coefficient of KIR2DL1 was found to be $\overline{D} = 0.23 \pm 0.06$ $\mu m^2 s^{-1}$ (Fig. 2C & D). The coefficient of determination $R^2$, which quantifies the "goodness of fit" of a predictive model, had a value of 0.90, indicating that the experimental data and the model simulations were highly correlated. It is important to note that due to cells rarely being perfect spheres and ruffling of the plasma membrane, our diffusion constants may underestimate the values of real diffusion coefficients. Figure 2E shows the distribution of values for the diffusion coefficients obtained by fitting each cell individually (see also Fig. S2 in the Supporting Material). The average $D_{av}$ of the values from individual fits coincides with the value $\overline{D}$ obtained by fitting the mean curve (Fig. 2D). Our value of the diffusion coefficient for KIR2DL1 is in the same order of magnitude as values obtained for other immune membrane receptors using other techniques, such as $0.12 \pm 0.01$ $\mu m^2 s^{-1}$ for the T cell receptor (22) and $0.067$ $\mu m^2 s^{-1}$ for the inhibitory NK cell receptor CD94/NKG2A complex (23).

One caveat of using tdEosFP is that since it is comprised of a dimer, it is a relatively large tag (approximately twice the size of GFP) that could potentially affect the mobility of KIR2DL1 at the plasma membrane. Therefore, to validate our experimental approach, we performed fluorescence recovery after photobleaching (FRAP) with a non-dimerizing tag. We performed FRAP experiments on unconjugated YTS cells transfected to express KIR2DL1 tagged to GFP. Using this method, we found that the fluorescence recovery curves obtained for KIR2DL1-GFP fit a free diffusion model and the average diffusion coefficient from $n = 11$ cells ($D_{av} = 0.22 \pm 0.09$ $\mu m^2 s^{-1}$) coincided with the coefficient obtained for KIR2DL1-tdEosFP (Fig. S3 in the Supporting Material). In future studies, an interesting option would be to use the newest version of EosFP, mEos2, a monomeric variant that folds efficiently at 37°C (24). Together, these data indicate that in unconjugated NK cells, i.e. in the absence of ligation, the KIR2DL1 receptor can diffuse freely within the plasma membrane of the cell (at the micrometer-scale imaged by confocal microscopy).

**Continuous recruitment of KIR2DL1 to the immune synapse**

To study how the dynamics of KIR2DL1 alters during the formation of an immune synapse, YTS/KIR2DL1-tdEosFP cells were imaged in the presence of 221/Cw4 target cells. When incubated together, the cells formed inhibitory immune synapses where KIR2DL1 accumulates at the site of contact between an effector cell and a target cell, as previously reported (6). To examine KIR diffusion, tdEosFP molecules were photoswitched in the region directly opposite the synapse (Fig. 3A & B).

Two alternative mathematical models could be used to describe the recruitment of receptors to the synapse. In the first model, an elastic potential introduces a force pointing towards the center of the synapse and proportional to the distance from the center, which traps the receptors within the synapse area (Fig. 3C, top). In the second model, slow diffusion inside the synapse was assumed due to crowding (Fig. 3C, bottom).

Interestingly, photoswitched KIR2DL1-tdEosFP molecules continuously trafficked to the immune synapse and accumulated there (Fig. 3A & D), and the red fluorescence detected in the region opposite the synapse steadily decreased (Fig. 3D). Both modeling approaches lead to an accumulation of the receptors inside the synapse in line with the experimental data (Fig. 3D, receptor density profiles shown in Fig. S4 in the Supporting Material). These results suggest that recruitment of KIR2DL1 to the immune synapse is a continuous and dynamic process.

Importantly, simulations run with the two models (Fig. 3C) show that a maximum coefficient of determination is achieved for a value of the diffusion coefficient outside the synapse close to the value obtained for the free diffusion case, i.e. with no synapse (Fig. 2E); the slow diffusion model provides a value $D = 0.23$ $\mu m^2 s^{-1}$, while the force model provides $D = 0.22$ $\mu m^2 s^{-1}$ (Fig. 3D). These results show that when an immune synapse has formed, KIR2DL1 receptors freely diffuse around the cell membrane before reaching the synapse. This is important because there has been considerable debate over whether or not KIRs are actively recruited to the synapse or not (6, 11). Crucially, these data argue against active recruitment, e.g. an ATP-dependent cytoskeletal process, for bringing KIR into the immune synapse.

**KIR2DL1 molecules remain trapped within the immune synapse**

To study the fate of KIR2DL1 molecules once at the immune synapse, YTS/KIR2DL1-tdEosFP cells were again imaged whilst in contact with 221/Cw4 cells, but only tdEosFP molecules present within the synapse were photoswitched from green to red fluorescence (Fig. 4A & B).

The fit with the two proposed models (Fig. 4C) is shown in Fig. 4D (note that each model was fitted simultaneously to the fluorescence data of Fig. 3D and 4D). Strikingly, the red fluorescence remained stable within the synapse (Fig. 4A & D). A slight decrease in signal over the time of imaging could be accounted for in the control experiments testing for the decrease in intensity due to photobleaching (Fig. S5 in the Supporting Material). The fluorescence detected in the opposite area remained very low and stable throughout (Fig. 4A & D). These results show that once the receptors have entered the immune synapse, they remain constrained, or 'trapped', within the immune synapse, suggesting that relatively little KIR2DL1 is rapidly internalized, degraded or diffuses away.

Figure 4D suggests that our two models with receptor confinement in the synapse are similarly good at describing the data - the overall quality of fit parameters $\chi^2$ and the coefficient of determination R show the force model to give a 10-15% better result (residual errors plotted in Fig. S6 in the Supporting Material). The best fit was obtained when a force parameter corresponding to 20 pN was chosen (see Discussion). In contrast, a model without receptor confinement in the synapse, i.e. one solely relying on ligand-receptor binding, was not able to explain the data (Fig. S7 in the Supporting Material).

**Trafficking of KIR2DL1 between synapses**

NK cells can form synapses with multiple target cells simultaneously, and indeed YTS/KIR2DL1-tdEosFP cells incubated with 221/Cw4 cells were occasionally seen to form two or more inhibitory synapses with target cells. This begs the question as to how KIR2DL1 molecules would behave in the context of multiple synapses with similar target cells. To address this, when NK cells with multiple (usually two) synapses were found, the tdEosFP molecules in one of the synapses were photoconverted and the cells were then imaged for ~20 minutes. In these experiments, there was a gradual accumulation of red fluorescence detected within the second synapse (Fig. 5A, B, C & Fig. S8 in the Supporting Material; observed in seven different experiments). Unexpectedly, these results establish that there is a transfer of proteins between synapses formed by the same effector cell with different target cells.

This finding led us to wonder about the mechanism behind this intersynaptic exchange of KIR2DL1 molecules. The most obvious hypothesis is that a small population of photoswitched molecules was able to leak out of the first synapse, undetected at the resolution of confocal microscopy, and was subsequently recruited to the second synapse by simple free diffusion. These molecules would either be too few to be detected or, more likely, too few to be differentiated from the red fluorescence background (see Fig. S5). To test this, we performed simulations using the same two models as above, but this time with two synapses, allowing for a small fraction of molecules to diffuse out of the first synapse (Fig. 5D). In both models, these assumptions were sufficient to reproduce the accumulation of molecules at a second synapse as seen in the experimental data and within the same timeframes (Fig. 5E and Fig. S9 in the Supporting Material). Thus, we conclude that although KIR2DL1 molecules are generally 'trapped' within an immune synapse, a small proportion of molecules is able to diffuse out and translocate into the other synapse.

**Discussion**

To investigate receptor mobility in the membrane, we applied a novel technique exploiting the photoswitchable properties of the tdEosFP protein. This fluorescent protein can be photoconverted from green to red fluorescence by UV light, allowing direct visualization of a specific population of fluorophores after photoswitching. We have shown that this technique can be successfully used to observe and quantitatively analyze the dynamics of membrane proteins in the setting of the NK cell immune synapse. Our study revealed new aspects of the behavior of the inhibitory receptor KIR2DL1 in the plasma membrane of NK cells. Specifically, we establish that KIR2DL1 molecules freely diffuse around the plasma membrane of unconjugated NK cells. In the presence of a target cell, an immune synapse is formed between the NK cell and the target cell, where KIR2DL1 molecules are continuously accumulated and to a large extent, remain trapped. Most surprisingly, despite the fact that receptors were found to be predominantly trapped within a single synapse, we observed that when multiple immune synapses were formed, KIR2DL1 molecules from one synapse were able to accumulate at a second synapse.

The method we applied here may prove to have several advantages over other commonly used techniques, since it allows the simultaneous observation of all the fluorescent molecules in the green and red fluorescence channels. It is also relatively easy to implement. The main technique that is currently used to study protein diffusion is fluorescence recovery after photobleaching (FRAP) (25, 26). Even though FRAP enables the determination of membrane protein diffusion coefficients, these are obtained by indirect measurements and this technique does not permit direct visualization of protein movement. Other techniques have also been developed to study protein mobility, such as fluorescence loss in photobleaching (FLIP), which is used to monitor the continuity of cellular compartments through continuous photobleaching of a specific small region (27–29) and fluorescence localization after photobleaching (FLAP), where the protein of interest carries two fluorophores, one of which is photobleached within localized regions of the cell (30). However, all these methods are based on photobleaching, which requires long and intense illumination of a specific region, and generally rely on indirect ways to track protein movements. Instead, photoswitching can provide a higher contrast between two subsets of molecules, allowing molecules to be tracked rapidly and more efficiently than photobleaching (31). By photoswitching fluorescent proteins, molecules can easily be directly visualized moving around the cell. This technique is particularly well suited to the study of the immune synapse as receptors can be photoswitched either within the synapse or elsewhere. Molecules can be photoconverted in one region and monitored as they move to another. In addition, when using this technique two regions of interest are monitored following a single photoswitching event (e.g. within the immune synapse and opposite the synapse), which allows the extraction of additional information through appropriate modeling. Although the use of photoactivatable and photoswitchable fluorescent proteins to study protein dynamics has been proposed and discussed (32, 33), this type of technique has not been extensively applied.

Regardless of whether or not an NK cell was in contact with a target cell, KIR2DL1 receptors freely diffused around the cell membrane. In contrast, the population of accumulated KIR2DL1 receptors at the immune synapse remained relatively stable, with little, if any, diffusion of molecules out of the synapse, indicating that KIR2DL1 molecules are effectively 'trapped' within the immune synapse. This suggests that immobilization of receptors at the site of contact, as described here, is a requirement for a sustained localized inhibitory signal within cells that are continuously poised to kill the target cells that they encounter (23). The kinetic parameters of KIR2DL1-HLA-C binding are known to be extremely rapid, with very fast association and dissociation rates (17). This is compatible with the idea that receptors trapped within the synapse may serve to enhance signaling through serial ligand binding.

We simulated the movement of receptors using two models, the first with receptors constrained within the synapse by a force and the second with receptor diffusion within the synapse significantly slower than outside the synapse. Comparison with experimental data showed that both models of receptor confinement can reproduce the integrated fluorescence data, although the force model provided a slightly better fit. This result suggests that a physical force may restrain KIR2DL1 receptors within the synapse. Our fitted force constant is consistent with typical forces generated by polymerization of a small number of actin filaments (1-10 pN) (34). The direction of the proposed force towards the center of the synapse may originate from retrograde actin flow, which has been observed previously (35). However, even in the slow diffusion model the reduced mobility of receptors might be caused by binding of receptors to the underlying actin meshwork in the synapse. This involvement of the actin cytoskeleton in regulating KIR2DL1 dynamics may be direct, such as in the picket fence model (36, 37), or indirect, through the attachment of other molecules (such as LFA-1) to the cytoskeleton, effectively causing trapping of KIR2DL1 molecules in the synapse. Alternatively, slow diffusion in the synapse may arise from the tethering of class I MHC proteins to the target cell actin cytoskeleton. Binding between KIR2DL1 and class I MHC proteins would link this complex to the target cell, which may influence the dynamics of KIR2DL1. There is evidence that the target cell cytoskeleton can influence the dynamics for NK cell synapses with dendritic cells (38), but perhaps not with other target cell types.

Consistent with actin-dependent receptor confinement in our two models, previous work has shown that the actin cytoskeleton can control the efficiency and speed at which KIR2DL1 receptors enrich at the immune synapse (11). When actin polymerization was inhibited, receptors have still been observed to accumulate at the immune synapse, albeit after longer times (11), suggesting that an intact actin cytoskeleton may be required for the rapid accumulation of KIR2DL1 receptors. Our quantitative analysis adds a new line of evidence for the hypothesis that receptors are physically trapped within the immune synapse, perhaps regulated by the actin cytoskeleton.

Both models are in line with the observed nanometer-scale organization of KIR2DL1 into nanoclusters within the synapse (14). KIR2DL1 receptor nanoclusters are both actin-dependent and slow down receptor diffusion. Indeed, KIR2DL1 has been observed in small clusters containing 15-30 proteins both in the membrane of resting NK cells and of cells in contact with a target cell (14). In resting cells, the formation of nanoclusters is consistent with the idea of membrane compartmentalization, with short-term confinement of proteins within domains and long-term 'hop diffusion' over the compartments (37). At the scale of microns – imaging whole cells (such as in our experiments) – this hopping effect is not visible but molecules effectively follow an apparent free diffusion as described here and observed elsewhere (39). We have previously shown that in cells forming an immune synapse, KIR2DL1 nanoclusters become smaller and denser and that these changes are dependent on an intact actin cytoskeleton (14), which is consistent with actin forces controlling the dynamics of KIR2DL1 behavior.

An observation that would not have been possible using photobleaching techniques to track receptors is that proteins from one synapse can traffic to a second immune synapse formed by the same NK cell with a second target cell. We showed that the movement of KIR2DL1-tdEosFP molecules from one synapse to another was rapid and in all cases occurred within a few minutes. It is surprising that although KIR2DL1 receptors are physically trapped within an immune synapse, they are able to traffic between synapses. The most likely explanation for this observation is that a small population of switched molecules – below the detection limit of confocal microscopy – diffuses out of the first synapse, diffuses in the cell membrane and gradually populates the second synapse. This hypothesis is supported by our simulations with two synapses. In the case of multiple synapses, it is possible that KIR2DL1 receptors that are already involved in signaling at one synapse and are then recruited to another synapse can enhance and accelerate the response to another target cell.

**Conclusion**

In summary, our use of mathematical models in combination with imaging a photoswitchable fluorescent protein has revealed new aspects of KIR2DL1 recruitment to the synapse including the surprising observation that proteins can traffic between synapses. Our analyses best fit a model in which KIR2DL1 freely diffuses around the cell membrane and then becomes trapped within the synapse which can be modeled as a synaptic force and reduced diffusion inside the synapse. The apparent receptor confinement within the synapse is likely related to the formation of a dense cytoskeletal meshwork at the immune synapse. These results give rise to new questions regarding the mechanisms regulating KIR2DL1 dynamics and recruitment, such as how receptors traffic between multiple synapses. Moreover, this study establishes how our method can be useful in studying protein dynamics in general, across a range of specific situations.

**Acknowledgements**

We thank Martin Spitaler and Stephen Rothery from the FILM imaging facility (Imperial College London) and members of our laboratories for discussion. We thank Shaun-Paul Cordoba for critical reading of the manuscript.

This research was funded by the Biotechnology and Biological Sciences Research Council (BBSRC funding to SVP), a Wolfson Royal Society Research Merit Award (to DMD), the Leverhulme Trust (Grant no. RPG-181 to GA and RGE) and a European Research Council Starting Grant (Grant no. 280492-PPHPI to RGE).

The authors declare no financial or other conflicts of interest.

**Supporting Citations**

References (40, 41) appear in the Supporting Material.

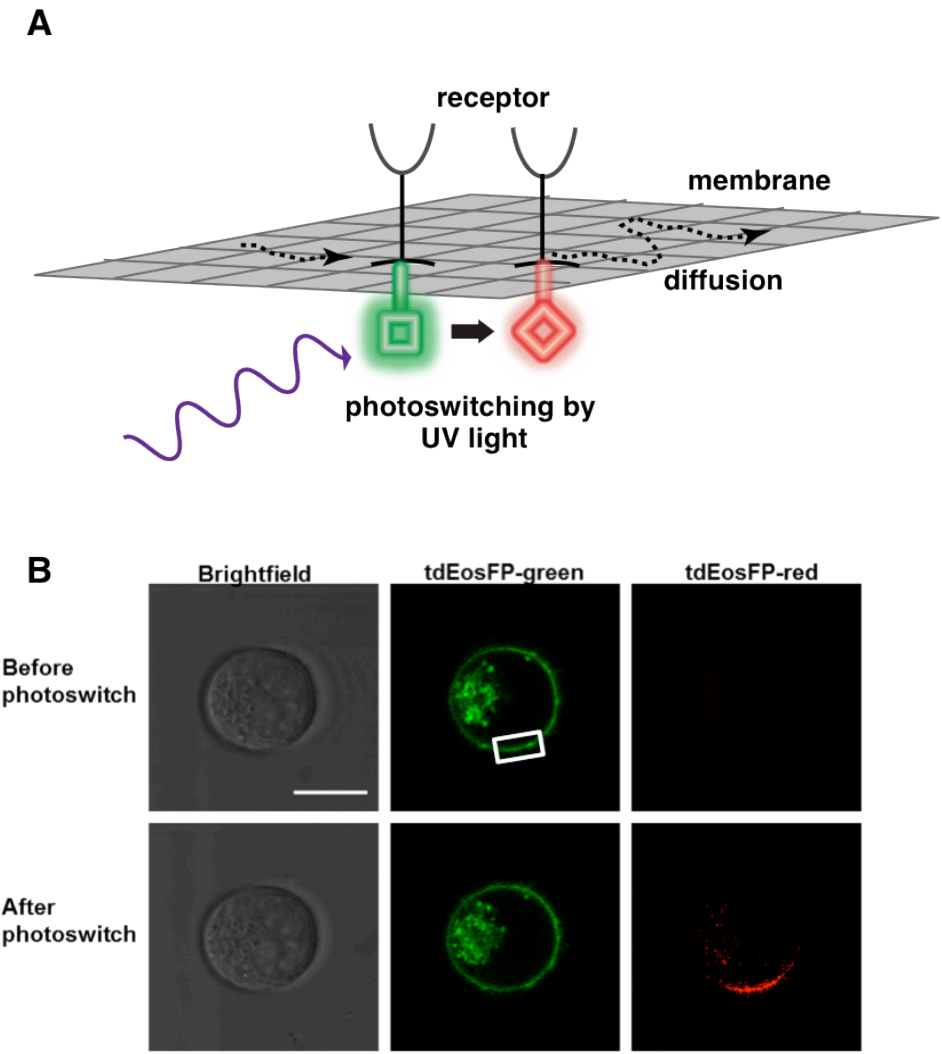

**Figure 1: Photoswitchable fluorescent protein as a tool for studying membrane protein dynamics.** **(A)** The photoswitchable fluorescent protein tdEosFP is capable of a photoconversion from green fluorescence (emission at 516 nm) to red fluorescence (581 nm) upon irradiation with UV light (~ 390 nm). By photoswitching tagged receptors only in limited areas, it is possible to follow their subsequent diffusion over the cell membrane against the green background. **(B)** YTS/KIR2DL1-tdEosFP cells were imaged by confocal microscopy. A specific region of the cell membrane was selected and targeted with UV light. The tdEosFP molecules within this region were seen to switch from green to red fluorescence. Images show a YTS/KIR2DL1-tdEosFP cell before and after photoswitching ($t$ = 20 s after photoswitching). Scale bar represents 10 μm.

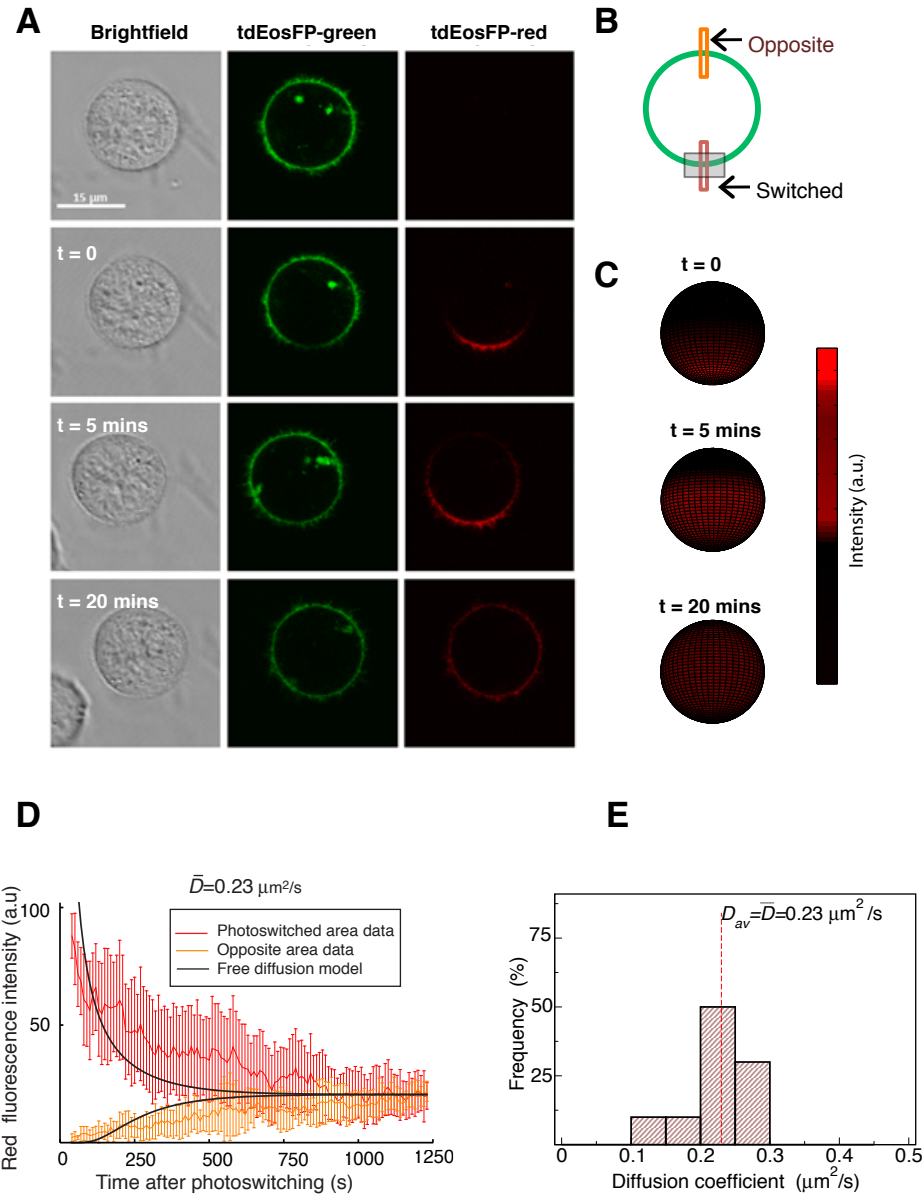

**Figure 2: KIR2DL1 dynamics fit a free diffusion model in the NK cell plasma membrane.** **(A)** Representative images of a YTS/KIR2DL1-tdEosFP cell, where a specific region of the cell membrane was targeted with UV light to photoconvert the tdEosFP molecules within this region ($t = 0$). The cell was subsequently imaged by confocal microscopy for 20 minutes and KIR2DL1 can be seen to diffuse around the cell. Scale bar represents 15 μm. **(B)** Schematic diagram showing the method used for measuring the diffusion of the receptor. The red fluorescence was measured in the photoswitched region and in the region directly opposite, as indicated by the boxes. The gray rectangle represents the region that was photoswitched. **(C)** 3D plotting on a sphere of the time evolution of the photoswitched receptor density according to the mathematical model (see Eq. 3 in main text) using the diffusion coefficients obtained by fitting to data ($D = 0.23$ μm$^2$s$^{-1}$). **(D)** Red fluorescence over time integrated over the photoswitched area (thin red line) and opposite area (thin orange line) with corresponding fits from the mathematical model of Eq. 3 in main text (thick red and orange lines). Data are taken from $n = 10$ cells and 3 independent

experiments. A threshold was applied to the images to remove background fluorescence. Error bars represent ± SEM. The diffusion coefficient of KIR2DL1, $D$, was determined by fitting the spherical diffusion equation to the experimental average data (red line). Fitting the mathematical modeling to the mean fluorescence intensity from the experiments leads to the value $\bar{D}$ = 0.23 µm$^2$s$^{-1}$ for the diffusion coefficient. **(E)** Distribution of the diffusion coefficient values obtained by fitting the mathematical model to each set of data of the *n* = 10 cells separately (see Fig. S2 in the Supporting Material). The average value coincides with the value obtained from fitting the mean curve (see (D)) and the variance around the average value is 0.06 µm$^2$s$^{-1}$.

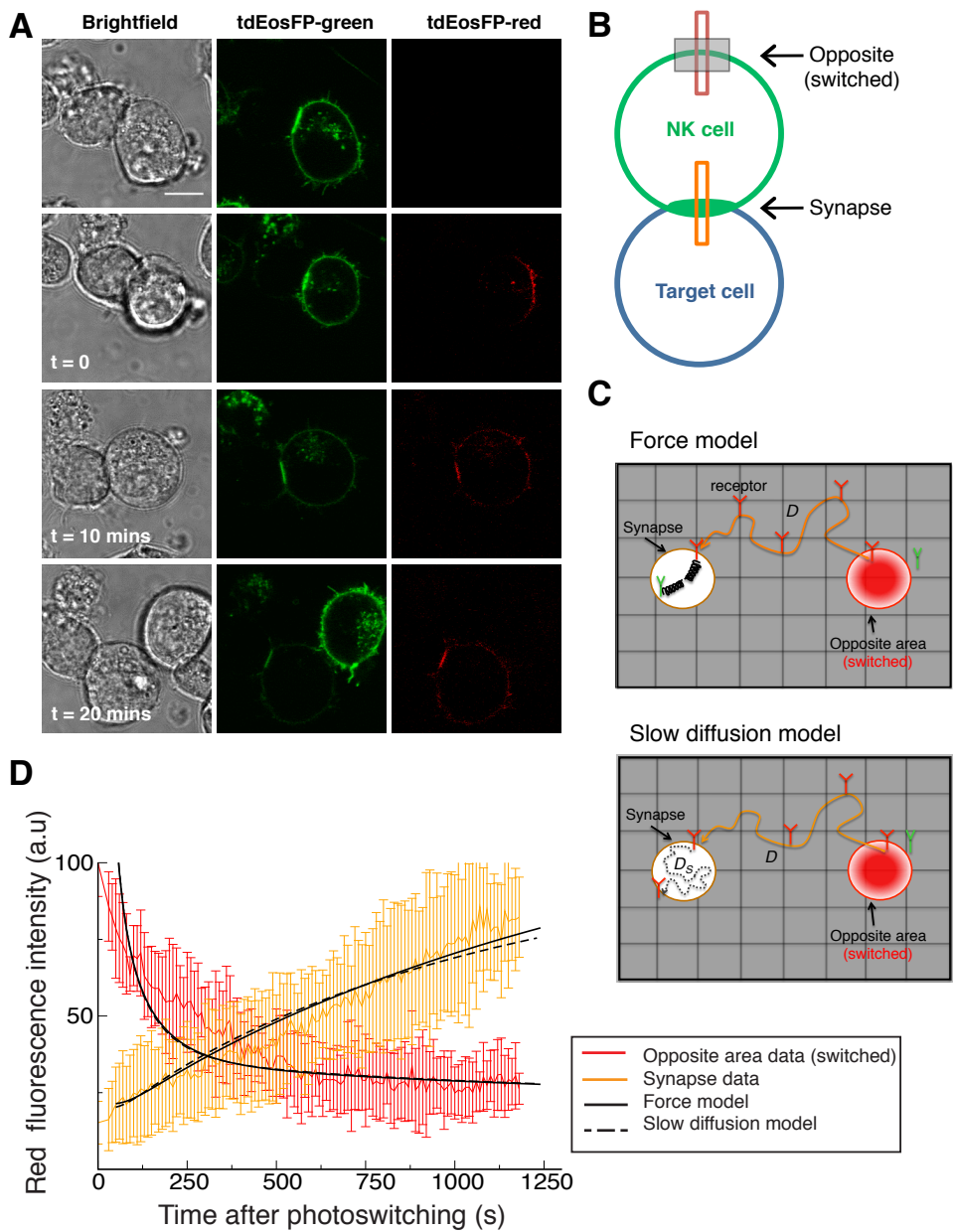

**Figure 3: KIR2DL1 is continuously trafficked to the immune synapse. (A)** Representative images of a conjugate formed between a YTS/KIR2DL1-tdEosFP cell and a 221/Cw4 cell, where the region of the cell membrane directly opposite the synapse was targeted with UV light to photoswitch the tdEosFP molecules within this region. The cells were imaged by confocal microscopy for 20 minutes and KIR2DL1 movement was monitored. Scale bar represents 10 µm. **(B)** Schematic diagram showing the strategy for monitoring the movement of the receptor. The red fluorescence was measured in the photoswitched region (opposite the synapse) and in the synapse region, as indicated by the boxes. The gray rectangle represents the region that was photoswitched. **(C)** Graphical representation of the two proposed models for receptor diffusion in the presence of a target cell: (top) Receptor diffusion on the cell membrane is modeled on a 2D grid where receptors diffuse freely outside the synapse and are trapped by a force inside the synapse. (bottom) Receptor diffusion is modeled with fast free diffusion outside the synapse and slow diffusion inside the synapse. Photoswitched area and receptors are marked in red. **(D)** Fluorescence intensity data observed after photoswitching receptors in an area opposite to the synapse and detected at the synapse (orange line) and opposite (red line) from $n = 10$ cells. Fits of the force model (solid black lines) and the slow diffusion model (dashed black lines) to the data (after rescaling to convert receptor density to fluorescence intensity). The best fit is obtained with a $\chi^2 = 91$ for the force model when the value of the force is taken as 20 pN and the value of the diffusion coefficient outside the synapse is $D = 0.22$ µm$^2$s$^{-1}$. For the slow diffusion model, the best fit is obtained with a $\chi^2 = 100$ for $D = 0.23$ µm$^2$s$^{-1}$ and $D_S = D/100$. Both fits have $\chi^2$ values corresponding to a probability < 0.5% of observing smaller value ($\chi^2$-test with

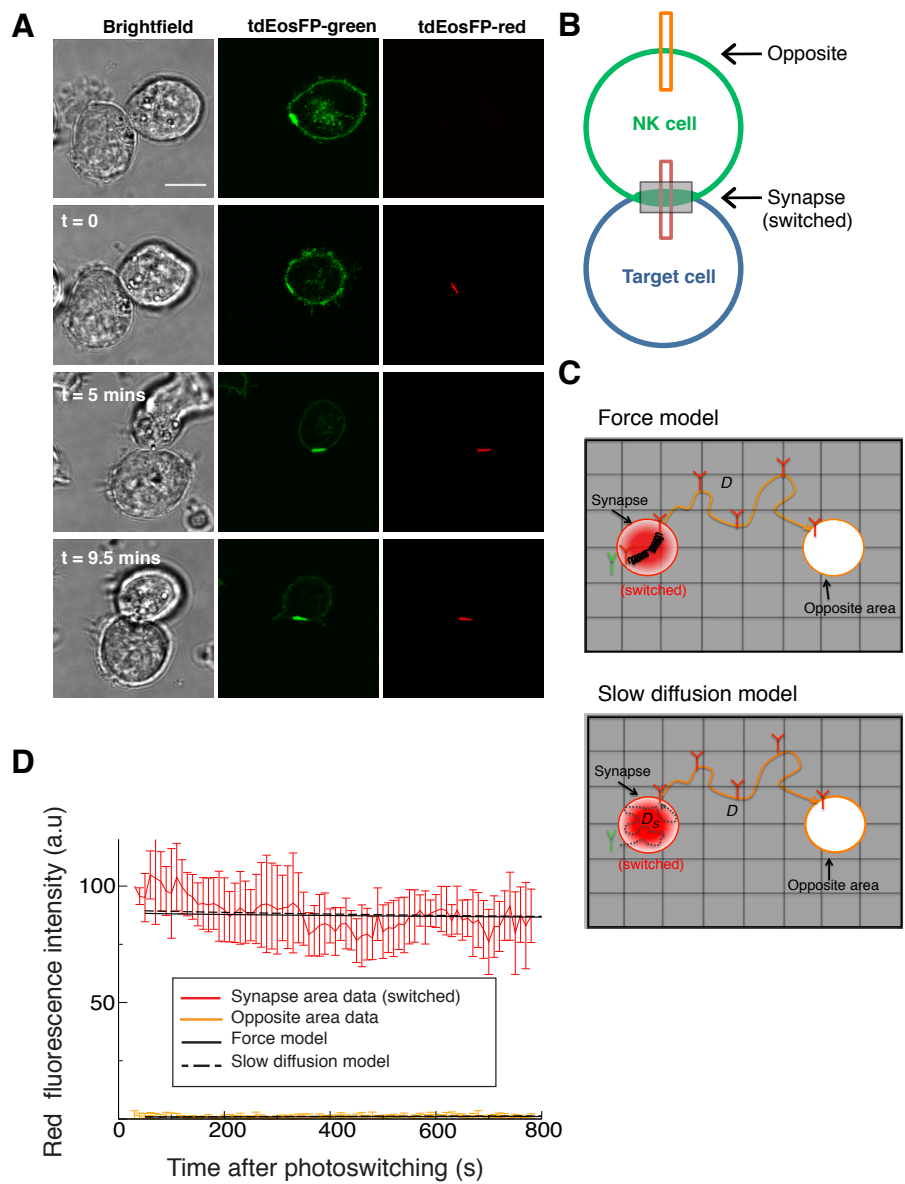

**Figure 4: KIR2DL1 molecules remain trapped at the immune synapse. (A)** Representative images of a conjugate formed between a YTS/KIR2DL1-tdEosFP cell and a 221/Cw4 cell, where the synapse region was targeted with UV light to photoswitch the tdEosFP molecules within the synapse. The cells were imaged by confocal microscopy for 10-15 minutes and KIR2DL1 movement was monitored. Scale bar represents 10 μm. **(B)** Schematic diagram showing the strategy for monitoring the movement of the receptor. The red fluorescence was measured in the photoswitched region (the synapse) and in the opposite region, as indicated by the boxes. The gray rectangle represents the region that was photoswitched. **(C)** Graphical representation of the proposed models, similar to Fig. 3C but for photoswitching inside the synapse: (top) force model, (bottom) slow diffusion model. **(D)** Fluorescence intensity data observed after photoswitching receptors inside the synapse. Fits of the force model (solid black lines) and the slow diffusion model (dashed black lines) to the data. Fitting with the models was performed on this and the data from Fig. 3D simultaneously (same fitting parameters, including diffusion coefficients and force parameter).

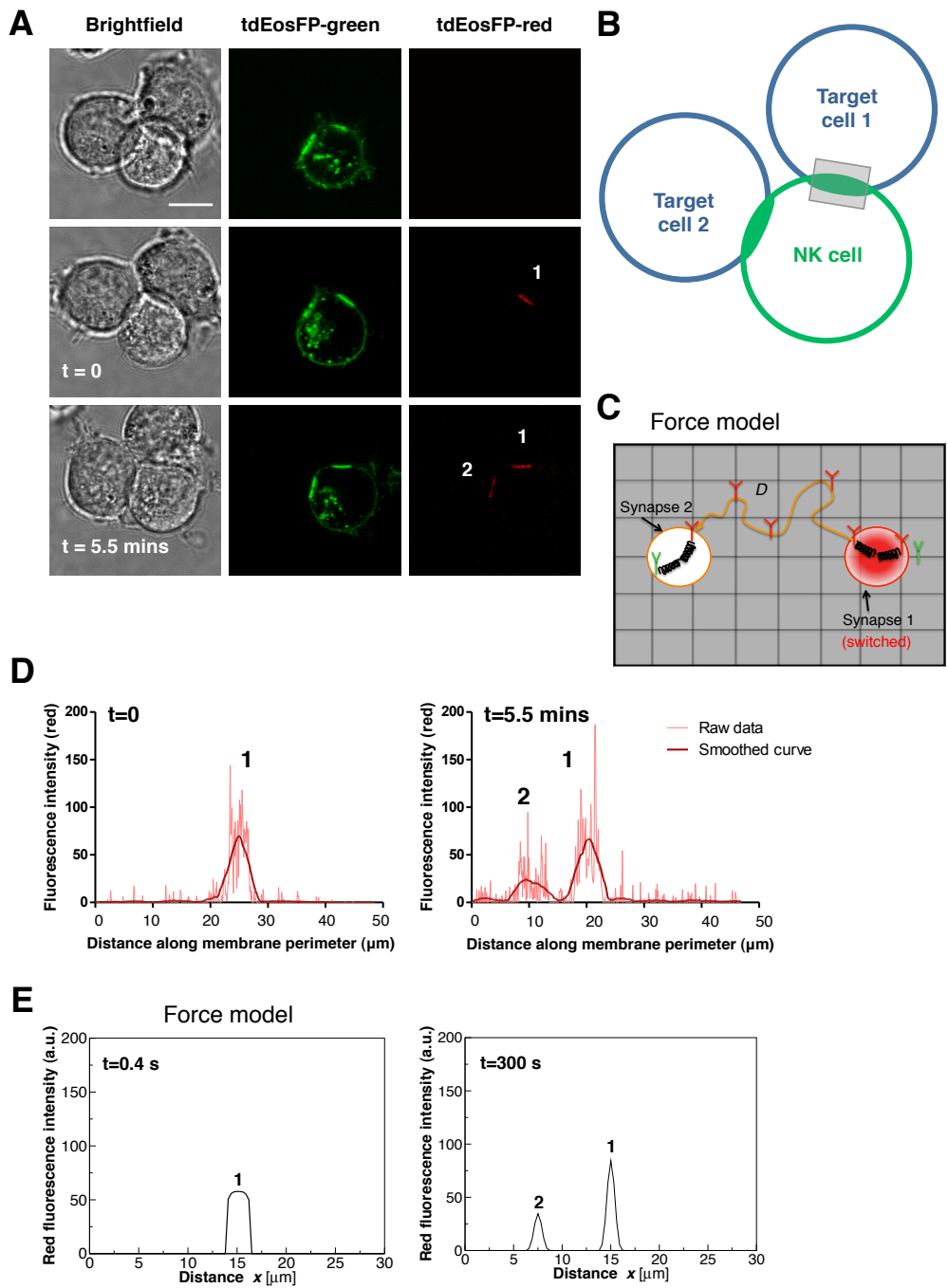

**Figure 5: KIR2DL1 can traffic between synapses. (A)** Representative images of a YTS/KIR2DL1-tdEosFP cell forming inhibitory immune synapses with two 221/Cw4 cells, where one of two synapses was targeted with UV light to photoconvert the tdEosFP molecules within this synapse. The cells were imaged by confocal microscopy for 10-15 minutes and KIR2DL1 movement was monitored. Scale bar represents 10 µm. **(B)** Schematic diagram showing the strategy for monitoring the movement of the receptor when two target cells form immune synapses with the same NK cell. One of the two synapses was photoswitched, as indicated by the gray rectangle. **(C)** Schematic of the force model applied to the case of two synapses. Two centers of force exist in this configuration, corresponding to each synapse. **(D)** The red fluorescence was measured around the periphery of the NK cell, just after photoswitching ($t = 0$) where one fluorescence peak is visible (corresponding to the photoswitched synapse), and again later on ($t = 5.5$ min) where a second fluorescence peak is appearing, showing KIR2DL1 accumulation at the second synapse. Data is representative of $n = 7$. **(E)** Results from simulations performed with the force model of Eqs. 4 in the case of two synapses located at a distance of 7.5 μm from each other, shown for times $t = 0.4$ s and $t = 5$ min. The parameters used for the force strength $K_s$ and the diffusion coefficient were the same as obtained from fitting experiments in Figs. 3D and 4D. Leaking between the synapses takes place on the same timescale as observed in experiments. For the implementation of the slow diffusion model in the case of two synapses and corresponding results, see Fig. S9 in the Supporting Material.